\documentstyle[12pt]{article}

\begin{document}

\LaTeX{}\bigskip\ \bigskip\ 

\begin{center}
The equation of state of a Bose gas: some analytical results\bigskip\bigskip%
\ \ 

Vladan Celebonovic\medskip\ 

Institute of Physics,Pregrevica 118,11080 Zemun,Yugoslavia\medskip\ 

vladan@phy.bg.ac.yu

vcelebonovic@sezampro.yu\bigskip\ 
\end{center}

Abstract:This is a short review of the main results on the equation of state
of a degenerate,non-relativistic Bose gas.Some known results are expressed
in a new form,and their possible applications in astrophysics and solid
state physics are pointed out.\bigskip\ \medskip\ 

\begin{center}
Introduction\medskip\ 
\end{center}

The aim of this letter is to review briefly some results concerning the
equation of state (EOS) of a non-relativistic,degenerate, Bose gas.The study
to be reported is a logical continuation of previous work concerning the
Fermi-Dirac integrals,and the EOS of a Fermi gas
(Celebonovic,1998,a,b).Apart from being useful pedagogically,this review
contains new expressions of some existing results.\medskip\ 

\begin{center}
Calculations\medskip\ 
\end{center}

It can be shown (for example Landau and Lifchitz,1976) that the number
density of a Bose gas is given by the following integral:

\begin{equation}
\label{(1)}n=\frac NV=\frac{gm^{3/2}}{2^{1/2}\pi ^2\hbar ^3}\int_0^\infty 
\frac{\epsilon ^{1/2}d\epsilon }{\exp [(\epsilon -\mu )/T]-1} 
\end{equation}

All the symbols have their usual meanings, $s$ is the particle spin,$g=2s+1$
and Boltzmann's constant has been set equal to 1.\newpage\ 

Introducing the change of variables $\frac \epsilon T=z$, it follows from
eq.(1) that

\begin{equation}
\label{(2)}n=\frac{g(mT)^{3/2}}{2^{1/2}\pi ^2\hbar ^3}\int_0^\infty \frac{
\sqrt{z}dz}{\exp [z-\frac \mu T]-1} 
\end{equation}
\medskip\ The function under the integral can be transformed as follows

\begin{equation}
\label{(3)}\frac{\sqrt{z}}{\exp \left[ z-\frac \mu T\right] -1}=\frac{\sqrt{z%
}}{\exp \left[ z-\frac \mu T\right] (1-\exp \left[ \frac \mu T-z\right] )} 
\end{equation}
\smallskip\ Developing eq.(3) into series,one gets the expression for the
integral in eq.(2)

\begin{equation}
\label{(4)}I_B=\int_0^\infty \frac{\sqrt{z}dz}{\exp \left[ z-\frac \mu
T\right] -1}=\sum_{l=0}^\infty \int_0^\infty \sqrt{z}\exp [\left( l+1\right)
(\frac \mu T-z)]dz 
\end{equation}
\smallskip\ 

which after some algebra can be transformed into the following final form

\begin{equation}
\label{(5)}I_B=\sum_{l=0}^\infty \exp \left[ (l+1)\frac \mu T\right]
\int_0^\infty \exp \left[ -(l+1)z\right] dz=\sum_{l=0}^\infty \exp
[(l+1)\frac \mu T]\frac{\sqrt{\pi }}2(l+1)^{-3/2} 
\end{equation}
\smallskip\ Generalizing the reasoning which led to eq.(5),it can be shown
that\medskip\ 

\begin{equation}
\label{(6)}I_n=\int_0^\infty \frac{z^ndz}{\exp \left[ z-\frac \mu T\right] -1%
}=\sum_{l=0}^\infty \exp [\left( l+1\right) \frac \mu T]\frac{\Gamma (n+1)}{%
(l+1)^{(n+1)}} 
\end{equation}
\medskip\ 

where $\Gamma $ denotes the gamma function. Inserting eq.(5) into eq.(1),one
gets the following form of the EOS of a Bose gas

\begin{equation}
\label{(7)}n=\frac{g(mT)^{3/2}}{2^{1/2}\pi ^2\hbar ^3}\sum_{l=0}^\infty 
\frac{\sqrt{\pi }}2\exp \left[ (l+1)\frac \mu T\right] (l+1)^{-3/2} 
\end{equation}
\newpage This EOS relates the chemical potential, temperature and number
density of a Bose gas.Inserting $\mu =0$ in eq.(7),one can determine the
temperature $T_B$ of Bose condensation. It thus turns out that

\begin{equation}
\label{(8)}T_B=\frac{2\hbar ^2}{\pi m}\left( \frac n{g\zeta (3/2)}\right)
^{2/3} 
\end{equation}
\smallskip\ where $\zeta $$(3/2)$ denotes Riemann's zeta function.

Usually,the EOS of a system is a relationship between its pressure,volume
and energy.It is known from general statistical physics ( such as Landau and
Lifchitz,1976) that for a Bose gas the EOS has the form

\begin{equation}
\label{(9)}pV=\frac 23E 
\end{equation}

\smallskip\ where V is the volume and E the energy of the system.The
expression for the energy of a Bose gas contains an integral of the form
given by eq.(6) with $n=3/2$ and the same prefactor as the right side of
eq.(1). Using eqs.(2),\{6) and (9),the following final result is obtained
for the EOS of a Bose gas in the region $T\succ T_{B.}$

\begin{equation}
\label{(10)}p=\frac{2^{1/2}g}{3\pi ^2}(\frac{mT}\hbar
)^{3/2}\sum_{l=0}^\infty \exp \left[ \left( l+1\right) \frac \mu T\right] 
\frac{\Gamma (5/2)}{\left( l+1\right) ^{5/2}} 
\end{equation}

\smallskip\ In the domain $T\prec T_B$,for which $\mu =0$ one gets that

\begin{center}
\begin{equation}
\label{(11)}p=\frac{2^{1/2}g}{3\pi ^2}(\frac{mT}\hbar )^{3/2}\Gamma
(5/2)\zeta (5/2) 
\end{equation}
\bigskip\ Discussion
\end{center}

In the preceeding section we have derived several analytical expressions for
the EOS of a Bose gas,both in the region of the existence of the Bose
condensation and outside it.The derivation of these expressions (and even
more their extension to the $T=0K$ case ) was an interesting problem in
itself.However,much more interesting are the possibilities for applications
of the results obtained here in studies of various systems occuring in
astrophysics and physics.\newpage\ 

Bose gas occurs in astrophysics in interesting and varied situations,which
range from the early universe to the interior of the giant planets.In
cosmology,Bose gases occur during the reheating after inflation.In a recent
paper (Khlebnikov and Tkachev, 1999) a nonequilibrium Bose gas with an
attractive interaction between particles was studied. It was shown there
that the system evolves into a state that contains drops of the
Bose-Einstein condensate.Could it be the possible explanation of the
formation of primordial clums out of which later clusters of galaxies were
formed?In planetary interiors ,of course depending on the chemical
composition,Bose gas can occur in some cases.Closely related is the problem
of the isolator $\rightarrow $ metal transition in planetary interiors.The
point here is that the transition occurs at a pressure which depends on the
chemical composition and the form of the EOS ( Stevenson,1998 for details on
this problem) .

Important applications of Bose gas thery are abundant in the theory of
superconductivity.It is known in ordinary superconductors that pairs pf
charge carriers form a superfluid Bose condensate.The mechanism responsible
for high temperature superconductivity has not yet been discovered
(Marston,1999),but it is almost certain that some form of a Bose gas will
also occur.The net conclusion from this short list of examples of
applicability testifies that studuing the EOS formalism is far from being a
mathematical ''tour de force'',but that quite to the contrary,it paves the
way to imoprtant astrophysical and physical applications.\medskip\ 

References\bigskip\ 

Celebonovic,V,.: 1998a, Publ.Astron.Obs.Belgrade,{\bf 60},16

Celebonovic,V.: 1998b, Publ.Astron.Obs.Belgrade,{\bf 61},75

Khlebnikov,S.and Tkachev,I.: 1999,preprint CERN-TH/99-21.

Landau,L.D. and Lifchitz,E.M.: 1976,Statisticheskaya fizika,Vol.I,Nauka
Publ.House,Moscow.

Marston,J.B.: 1999,preprint cond-mat/9904437 . 

Stevenson,D.J.:1998,J.Phys.:Condens.Matter., {\bf 10},11227.\ 

\end{document}